\documentclass[aps,prb,twocolumn,superscriptaddress]{revtex4-2}
\usepackage{xcolor}
\usepackage{amsmath}
\usepackage{graphicx}
\newcommand{\av}[1]{\left\langle #1 \right\rangle}
\newcommand{\up}{\uparrow}
\newcommand{\dn}{\downarrow}

\begin{document}

\title{Charge correlation, doublon-holon binding and screening in the doped Hubbard model}

\author{Edin Kapetanovi\'{c}}
\affiliation{I. Institute of Theoretical Physics,
	Universit\"at Hamburg,
	D-22607 Hamburg,
	Germany}
\author{Guglielmo Nicola Gigante}
\affiliation{NanoLund and Division of Mathematical Physics, Department of Physics, Lund University, Professorsgatan 1, Lund, Sweden}
\author{Malte Sch\"{u}ler}
\affiliation{Institut f\"{u}r Theoretische Physik, Universit\"{a}t Bremen, Otto-Hahn-Allee 1, D-28359 Bremen}
\affiliation{Bremen Center for Computational Materials Science, Universit\"{a}t Bremen, Am Fallturm 1a, D-28359 Bremen}
\author{Tim O. Wehling}
\affiliation{I. Institute of Theoretical Physics,
	Universit\"at Hamburg,
	D-22607 Hamburg,
	Germany}
\affiliation{The Hamburg Centre for Ultrafast Imaging,
	Luruper Chaussee 149,
	D-22761 Hamburg,
	Germany}

\author{Erik van Loon}
\affiliation{NanoLund and Division of Mathematical Physics, Department of Physics, Lund University, Professorsgatan 1, Lund, Sweden}

\date{\today}

\begin{abstract}
Electronic correlations arise from the competition between the electrons' kinetic and Coulomb interaction energy and give rise to a rich phase diagram and many emergent quasiparticles. The binding of doubly-occupied and empty sites into a doublon-holon exciton is an example of this in the Hubbard model. Unlike traditional excitons in semiconductors, in the Hubbard model it is the kinetic energy which provides the binding energy. Upon doping, we find the emergence of exciton complexes, such as a holon-doublon-holon trion. The appearance of these low-lying collective excitations make screening more effective in the doped system. As a result, Hubbard-based modelling of correlated materials should use different values of $U$ for the doped system and the insulating parent compound, which we illustrate using the cuprates as an example.
\end{abstract}

\maketitle



\section{Introduction}

Although the Coulomb interaction between electrons is fundamentally repulsive, its final effect can be attractive and bound many-electron quasiparticles and phase transitions resulting from the Coulomb interaction are at the heart of condensed matter physics. In magnetism, the combination of the Coulomb interaction and the antisymmetry of the wavefunction generates the exchange interactions responsible for magnetic phases~\cite{Szilva23}. In semiconductors, working with valence band holes changes the sign of the Coulomb interaction and explains the binding of excitons and more complex emergent quasiparticles such as trions~\cite{Lampert58}. The spatial structure of the Coulomb interaction leads to charge-density waves~\cite{Tosatti74,Hansmann13} and Wigner crystallization~\cite{Wigner34}, where a non-uniform ground state minimizes the Coulomb energy. On the other hand, the dynamic structure of the phonon-screened Coulomb interaction leads to the effective attraction between Cooper pairs in BCS superconductivity~\cite{SchriefferBook}. Finally, for unconventional superconductors, it is postulated that collective electronic excitations take over the role of phonons and provide the pairing glue~\cite{Norman11}. 

The essential elements of several Coulomb-driven emergent phenomena show up in the Hubbard model~\cite{Hubbard1,Hubbard2,Hubbard3} and its extensions, for example metal-insulator transitions~\cite{Imada98}, magnetism~\cite{KanamoriHubbardModel1963,GutzwillerHubbardModel1963,SchaeferFateOfMottHubbardTransition}, charge-density waves~\cite{Zhang89} and unconventional superconductivity~\cite{JiangDevereauxSuperconductivity}. Although the model is generically hard to solve~\cite{Troyer05}, there has been tremendous numerical progress in recent years~\cite{JiangDevereauxSuperconductivity,SimonsPureHubbardModel,SchaeferMultiMethod}, especially when it comes to the determination of ground state properties of the square lattice Hubbard model. It has an antiferromagnetic ground state at zero temperature and half-filling (one electron per site), which is destroyed by Mermin-Wagner fluctuations at $T>0$~\cite{SchaeferFateOfMottHubbardTransition}, but with very strong antiferromagnetic correlations still present.


This antiferromagnetic ground state at half-filling forms a starting point for understanding the effect of doping. If the repulsive interaction $U$ is large, doping introduces holes into an antiferromagnetic background, while double occupancy remains forbidden. One of the insights coming from the $t$-$J$ model~\cite{AndersontJ} is that these holes can propagate as pairs of empty sites (holon-pairs) without disturbing the antiferromagnetic ordering. This pair binding has been suggested as a possible mechanism for unconventional superconductivity. When $U$ is similar to the bandwidth, double occupancy is suppressed but not completely forbidden (e.g., 5\% double occupancy at $U/t=8$, half-filling and low temperature~\cite{Leblanc15}) and the $t$-$J$ approximation is no longer valid. The presence of doublons (doubly occupied sites) and their possible contribution to binding should now be considered. Do doublons repel or attract holons? This question of doublon-holon binding becomes more important as doping increases, since more and more holons are present in the ground state. Apart from doublon-holon pairs, it is even possible to form higher-order exciton complexes. 

With hole doping, the empty sites also tend to order spatially, e.g., in the form of stripes~\cite{PoilblancStripes,ZhengStripes17}. The Hubbard interaction is not sensitive to the spatial structure of the holons, so it must be the kinetic energy that drives this ordering.
At the same time, the appearance of these phases with non-homogeneous electron density raises questions about the applicability of the Hubbard model, i.e., the neglect of nonlocal Coulomb interactions. For a non-uniform ordered phase, these explicitly contribute to the total energy. Zheng et al.~\cite{ZhengStripes17} estimate an energy difference of $0.01t$ between vertical stripes and uniform superconducting states in the Hubbard model, and hole densities that differ by roughly 0.1 electron/site. With $\Delta E \approx V \, \Delta n_i \,\Delta n_j$, and given that $V\approx 2t$ in the cuprates~\cite{HirayamaCuprateSCHamiltonians}, the potential energy difference between charge-ordered and uniform phases due to the nonlocal Coulomb interaction has the same order of magnitude and cannot simply be ignored. This has motivated the study of the effect of nonlocal interactions on charge ordering~\cite{GullCDW2DHubbard,BariCDW,Zhang2DExtendedHubbard} and superconductivity~\cite{ExtendedHubbardSuperconductivity1,ExtendedHubbardSuperconductivity2,ExtendedHubbardSuperconductivity3}.

The impact of nonlocal Coulomb interactions can already be assessed based on knowledge of the correlation functions in the Hubbard model, using a variational principle~\cite{OptHubbardModelsMalte}. Essentially, the charge correlations of the Hubbard model show how the nonlocal Coulomb interactions effectively screens the Hubbard interaction. This approach has been applied extensively to half-filled systems~\cite{OptHubbardModelsMalte,MalteFirstOrderMITScreening,MalteThermodynamicsMITExtendedHubbard,EdinHubbardHeisenberg}, but the doped case has received less attention until now, due to the numerical difficulty caused by the Monte Carlo sign problem. 

\begin{figure}
	\centering
	\includegraphics[width=\columnwidth]{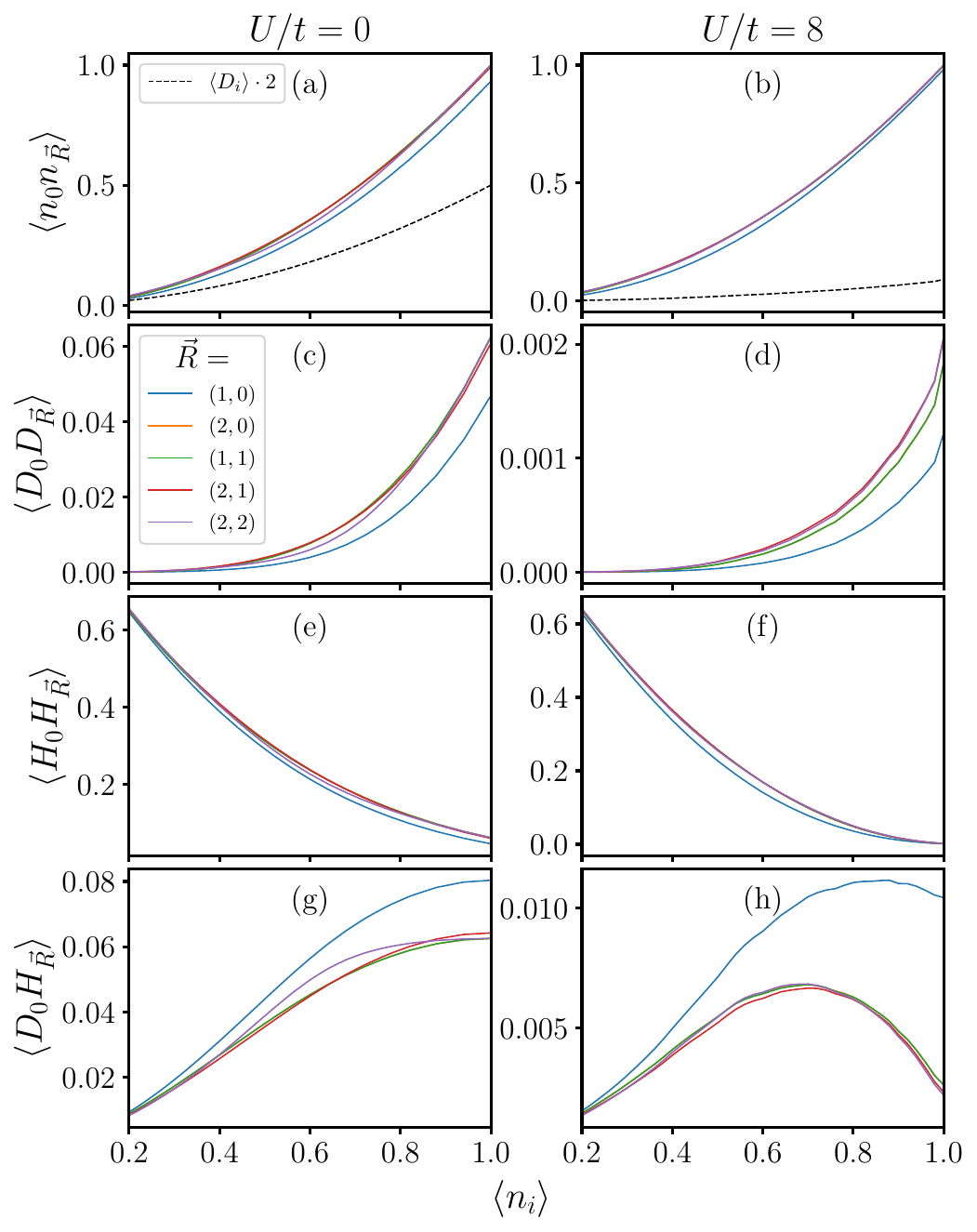}
	\caption{
 Spatial charge correlations in the Hubbard model, DQMC on a $4\times 4$ periodic cluster at $\beta t=2$. (a-b) Density-density, (c-d) doublon-doublon, (e-f) hole-hole (g-h) doublon-hole. Plots for intermediate values, i.e. $U/t = 2,4,6$ are provided in the Appendix (Fig \ref{fig:4particle_U_246}).
 }
 \label{fig:4particle}
\end{figure}

Here, we study the charge fluctuations in the doped Hubbard model using Quantum Monte Carlo. First, we quantify the presence of charge correlations as a function of doping and interaction strength. Then, we proceed with the spatial correlation between doublons and holes, to investigate if there are signs of exciton binding at the four-particle level. We focus on the regime up to $U \leq W$, where doublons are sufficiently present and where Quantum Monte Carlo studies of the doped system are numerically feasible. Finally, we study how screening is affected by doping, and the implications for the doping-dependence of the effective local interaction~\cite{OptHubbardModelsMalte} in correlated materials modelling. 

\section{Solving the Hubbard model}

We consider the Hubbard model on the square lattice, in the grand canonical ensemble, i.e.,
\begin{equation}
    H_\text{Hub} = - t \sum_{\langle i,j \rangle, \sigma} \left( c_{i\sigma}^\dagger c_{j\sigma}^{\vphantom{\dagger}} + h.c. \right) + U \sum_i n_{i\uparrow} n_{i\downarrow} - \mu \sum_i n_i.
    \label{eq:hubbard}
\end{equation}
Here, $c_{i\sigma}^{(\dagger)}$ denotes the annihilation (creation) operator for an electron on site $i$ with spin $\sigma$, $n_{i\sigma}$ is the corresponding occupation number operator, $n_i = n_{i\uparrow} + n_{i\downarrow}$ the total occupation number. The physical parameters are the nearest-neighbor hopping $t$, the on-site or Hubbard interaction $U$ and the chemical potential $\mu$. We consider the hole-doped case, i.e., $\av{n_i}\leq 1$ and $\mu\leq \mu_{1/2}= U/2$.

We study this model at finite temperature using Quantum Monte Carlo simulations, namely the Determinantal Quantum Monte Carlo method~\cite{DQMCIntroduction} (DQMC) as implemented in the QUEST code~\cite{QUESTLink} and the Dynamical Cluster Approximation~\citep{DCAHettler1998} (DCA) using a CT-AUX solver~\cite{CTAUXPaperEmanuel}, see Appendix~\ref{sec:computational} for further details.
The size of the simulation cell is an important limitation for both methods, but since the finite size effects are handled in different ways, both methods are complementary. We restrict ourselves to a 4x4 periodic square lattice within DQMC due to the severity of the Monte Carlo sign problem, and to an 8-site and 16-site dynamical cluster within DCA. The charge-correlation functions of interest can be obtained directly from the respective clusters. 

For both methods, raw Monte Carlo data is obtained on a dense grid in ($U$-$\mu$) space, and is then smoothed by a Savitzky-Golay filter \cite{SavitzkyGolay}. This filter mitigates the problem with noise when numerically evaluating derivatives, as needed later for the determination of effective Hubbard interactions. To facilitate the comparison between different values of $U$, the data is mapped from $\mu$ to $\av{n}$ in the figures. More details about the computations can be found in the Appendix~\ref{sec:ED}, which also contains Exact Diagonalization results obtained using EDLib~\cite{Iskakov18} as a reference.

\section{Charge fluctuations}

Figure~\ref{fig:4particle}(a-b) shows the density-density correlation $\av{n_0 n_{\vec{R}}}$ and the double occupancy (i.e., $\vec{R}=0$). Both increase with $\av{n}$, as expected. The main effect of $U$ is to suppress the double occupancy, while the other charge correlation functions are weakly enhanced: with fewer doubly occupied sites, there are also fewer empty sites, so the \textit{instantaneous} charge distribution is more homogeneous. Comparing the different values of $\vec{R}$ (colored lines) shows that the nearest-neighbor correlation function is always smaller than the ones further away. Avoiding charges on neighboring sites allows for more hopping processes and therefore lowers the kinetic energy.

Knowing that there are doubly-occupied sites, we continue with their spatial distribution. Figure~\ref{fig:4particle}(c-d) shows that doublons repel each other, since the doublon-doublon correlation function is smaller for neighboring sites than for sites further apart. This effect is already present at $U=0$, where $\av{D_i D_j}=\av{n_{i\sigma}n_{j\sigma}}^2$. Even with the overall suppression of the number of doubly occupied sites by $U$, doublons still repel and are unlikely to sit on neighboring sites, since that is bad for the kinetic energy. Similarly, Figure~\ref{fig:4particle}(e-f) shows that holons are less likely to occupy neighboring sites. In both cases, by increasing $U$, the charge localizes and correlations beyond nearest neighbor are weakly dependent on distance.

On the other hand, doublons and holons are more likely to occupy neighboring sites, as shown in Fig.~\ref{fig:4particle}(g-h). Again, this effect is already present at $U=0$, but it is enhanced by $U$, especially for $\av{n}>0.7$. In a real-space strong-coupling picture, putting doublons and holons next to each other maximizes the kinetic energy gain. In other words, virtual hopping of one electron to a neighboring site is the main mechanism creating doublons in a doped antiferromagnet (see Figure~\ref{pic:Mapping}), and the kinetic energy provides the binding for this doublon-holon exciton. Of the spatial correlation functions considered here, the doublon-holon binding is by far the strongest effect.

\section{Variational approach to nonlocal Coulomb interactions}

\begin{figure}[htp!]
	\centering
	\includegraphics[width=0.4\textwidth]{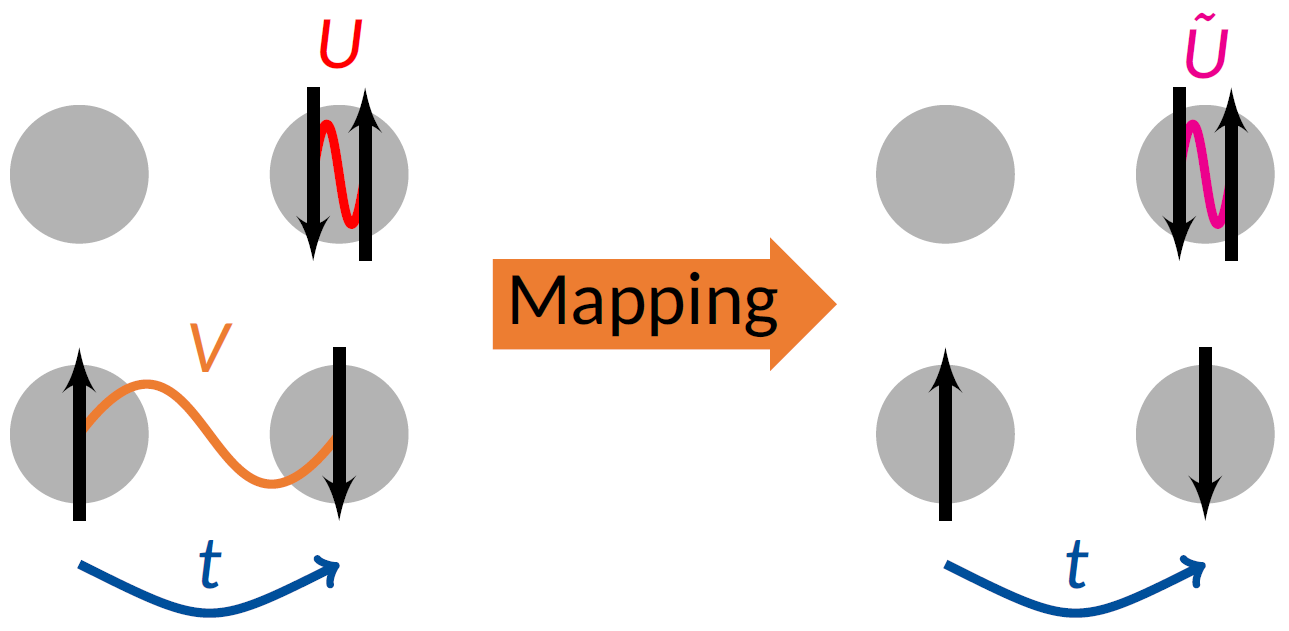}
 
  \vspace{0.5cm}
	\includegraphics[width=0.5\textwidth]{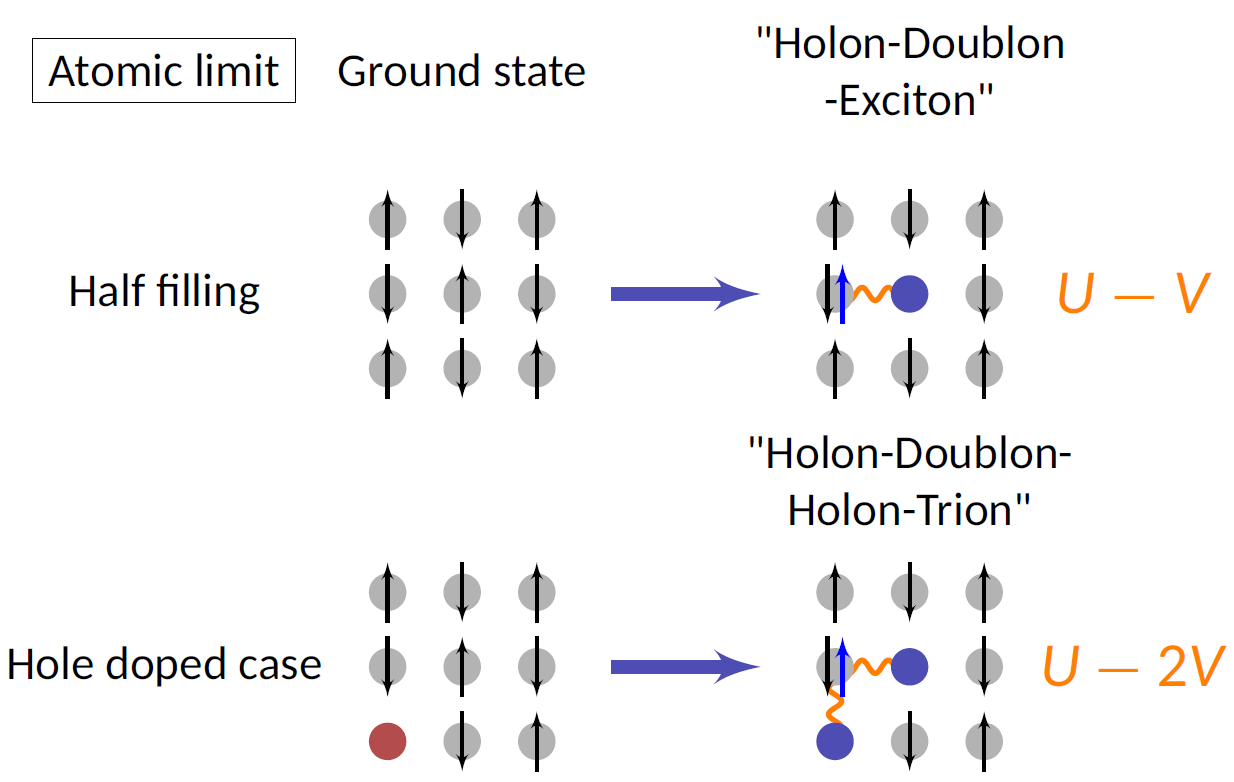}
	\caption{Extended Hubbard model can be mapped onto an effective model with local interactions only. The value of $\tilde{U}$, given by Eq.~\eqref{eq:Utilde} is based on the excitations in the model. In the antiferromagnetic ground state in the atomic limit, at half-filling the basic excitation is the creation of a holon-doublon exciton, while the formation of holon-doublon-holon trions is possible in the doped system.
 }
 \label{pic:Mapping}
\end{figure}


\begin{figure}[htp!]
	\centering
	\includegraphics[width=\columnwidth]{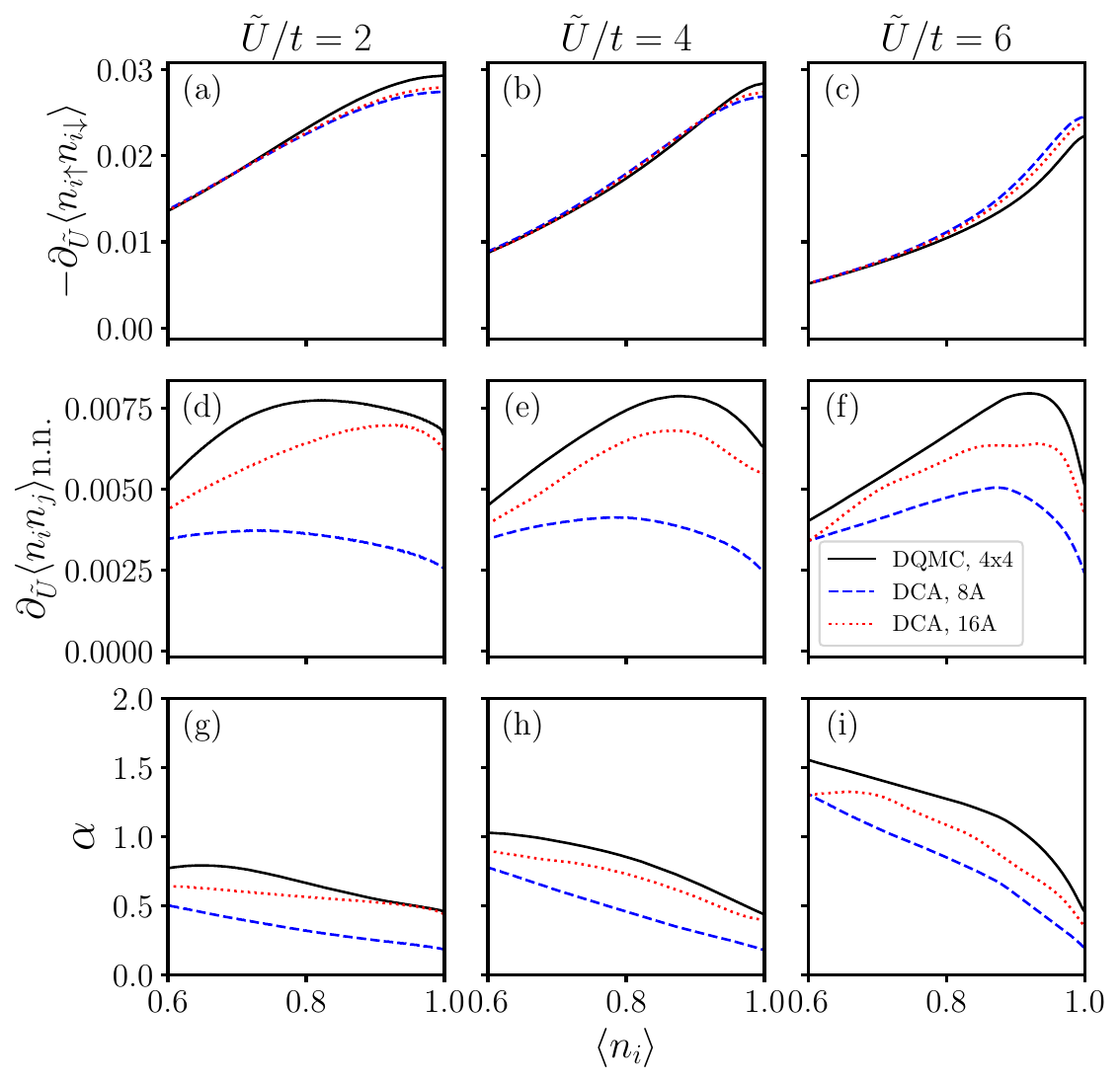}
	\caption{Screening factor $\alpha$ and the $\tilde{U}$-derivatives of the double occupancy and the next-neighbor density-density correlation vs. filling $\langle n_i \rangle$ at $\beta t = 2$, for different $\tilde{U}$. For the square lattice, DCA and DQMC lead to qualitatively similar results.}\label{pic:AlphaResults}
\end{figure}

\begin{figure}
	\centering
	\includegraphics[width=0.9\columnwidth]{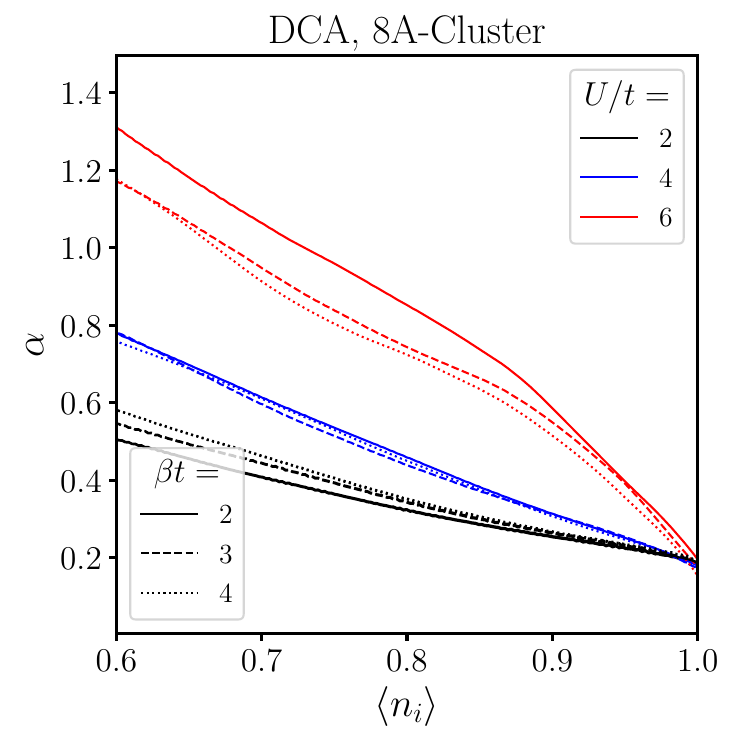}
 \caption{Screening factor $\alpha$ for different inverse temperatures $\beta$: $\beta t=2$ (full), $\beta t=3$ (dashed) and $\beta t=4$ (dotted), obtained using 8-Site Cluster (DCA).}
\label{pic:BetaDependence:alpha}
\end{figure}

Going beyond the Hubbard approximation, two-dimensional materials have nonlocal Coulomb interactions of substantial magnitude, and these directly affect the charge-correlation function. Here, we consider the extended Hubbard model with nearest-neighbor Coulomb interaction $V$,
\begin{equation*}
 H_\text{eHub} = H_\text{Hub}+ V \sum_{\langle i,j \rangle} n_i n_j \label{eq:extended}
\end{equation*}
Note that it is also possible to write the nonlocal interaction in terms of fluctuations away from half-filling, i.e., $V\sum_{\av{i,j}} (n_i-1)(n_j-1)$. This formulation is equivalent up to shifts in the chemical potential and total energy, and can simplify pictorial arguments substantially.

Using the Peierls-Feynman-Bogoliubov Variational principle~\cite{PeierlsVariationalPrinciple,FeynmanVariationalPrinciple,BogoliubovVariationalPrinciple}, it is possible to map~\cite{OptHubbardModelsMalte} the extended Hubbard model onto a regular Hubbard model with modified parameters $\tilde{U}$, $\tilde{\mu}$, which are variational and should be chosen so that the density operator of the effective Hamiltonian $\tilde{H}$ approximates the real density operator as well as possible, as illustrated in Fig.~\ref{pic:Mapping}. 


Although Quantum Monte Carlo works in the grand canonical ensemble, one usually adjusts the chemical potential to obtain a desired filling. Thus, for the variational approach, we choose $\tilde{\mu}$ as a function of $\tilde{U}$ in order to obtain a fixed filling ($\langle n_i \rangle_{\tilde{H}} = \textrm{const.}$), and $\tilde{U}$ remains as the sole variational parameter. Then, the effective local interaction $\tilde{U}$ is given by~\cite{OptHubbardModelsMalte,vanLoon16}:
\begin{align}
    \tilde{U} &= U - \alpha (\tilde{U}) ,V\label{eq:Utilde} \\
    \alpha (\tilde{U}) &= - \frac{Z}{2} \frac{\partial_{\tilde{U}}\langle n_i n_j \rangle_{\tilde{H}}}{\partial_{\tilde{U}} \langle n_{i\uparrow} n_{\downarrow}\rangle_{\tilde{H}}} .\label{eq:AlphaDefinition}
\end{align}
$\alpha$ is the \textit{screening factor} which determines how strongly the nonlocal Coulomb interaction effectively changes the local one. It is the central quantity in the variational approach and only depends on the properties of the effective Hubbard model, so it can be extracted from the available Quantum Monte Carlo data. 

The interpretation of the renormalized Hubbard interaction is that it relates the cost of creating a doublon-holon excitation in both models, taking into account the spatial correlation between them. The value of $\tilde{U}$ is chosen so that the ``typical'' doublon excitation has the same cost as in the extended Hubbard model Eq.~\eqref{eq:extended}. At half filling and strong coupling, $\tilde{U} = U - V$~\cite{OptHubbardModelsMalte}, i.e. $\alpha = 1$, corresponding to a strongly bound nearest-neighbor doublon-holon exciton, as shown in Fig.~\ref{pic:Mapping}. Previous studies have shown~\cite{vanLoon16} that $\alpha<1$ at smaller $U$ and half-filling, since the doublon-holon pair is more delocalized, as is also visible in Fig.~\ref{fig:4particle}(g-h).


For the doped system, Figure~\ref{pic:AlphaResults} shows the $\tilde{U}$-derivatives that make up the numerator and denominator of Eq.~\eqref{eq:AlphaDefinition} as well as the screening factor $\alpha$. The behaviour is qualitatively similar in DQMC and 8-site and 16-site DCA, showing that the observed mechanisms are robust with respect to finite-size errors. Importantly, the screening factor $\alpha$ \textit{increases} when doping the system away from half-filling, and $\alpha>1$ for a large range of filling at the largest shown value of $U$. 

In order to gain a physical understanding of the observed increase in screening upon doping, we go back to the atomic limit, as shown in Fig.~\ref{pic:Mapping}. Whereas half-filling has the nearest-neighbor doublon-holon exciton forming from a uniform background as the elementary excitation ($\alpha=1$), the doped system already has holes present in the ground state, so the created doublon-holon pair can bind to an existing holon and form a holon-doublon-holon trion. This costs energy $U-2V$, which is less than $U-V$ for a normal exciton, and which would lead to $\alpha\approx 2$ if it was the only relevant process. The observed $\alpha$ is an indicator of the relative statistical importance of holon-doublon excitons ($\alpha=1$) and higher-order holon-doublon exciton complexes ($\alpha=2$ for the trion). As the system is doped away from half-filling, the number of holes present in the ground state increases, and therefore also the probability of forming the compound quasiparticles, as visible in Fig~\ref{pic:AlphaResults}. The presence of these lower-lying excitations explains why the additional presence of holes leads to a more effective screening process. At very strong doping, this picture of a uniform antiferromagnetic background eventually breaks down, and we would expect $\alpha$ to decrease again since in the limit $\av{n}\rightarrow 0$ there are no electrons to do the screening. 

Since the variational principle is based on the free energy, the screening factor $\alpha$ depends on temperature via the correlation functions entering Eq.~\eqref{eq:AlphaDefinition}. The temperature-dependence of the double occupancy has been studied in detail~\cite{Leblanc15,MalteFirstOrderMITScreening,Sushchyev22,Roy24}.
In $\alpha$, this leads to visible differences between $T=t/2$ and $T=t/3$, but smaller differences going to $T=t/4$, as shown in the DCA results of Figure~\ref{pic:BetaDependence:alpha}. Thus, while the numerical sign problem prohibits us from performing Monte Carlo simulations of the doped model at low temperature, it is reasonable to assume that the effect of increased screening when doping the system away from half filling remains also at lower temperature. As a complementary method, we have performed exact diagonalization calculations at $T=0$, as shown in the Appendix.

\section{Conclusions and Outlook}

In conclusion, the presence of holes in the doped Hubbard model makes spatial charge correlations omnipresent, even at strong coupling. One of the main effects is the binding of doublon-holon excitons on neighboring sites, an effective non-local interaction between charged particles that arises from the kinetic energy in the Hubbard model. In terms of screening, the doped Hubbard model has an additional screening process in the form of holon-doublon-holon coupling, due to the holes in the antiferromagnetic background. This leads to screening factors $\alpha>1$ in the mapping to an effective Hubbard model, and thus to lower effective interactions. 

Concretely, it means that a lower value of $U$ should be used to model the doped cuprates than to model their half-filled parent compounds, if the modelling is done at the level of the Hubbard model, i.e., with non-local Coulomb interactions integrated out. Estimates of $t$, $U$ and $V$ for the cuprates~\cite{HirayamaCuprateSCHamiltonians} in a downfolded one-band model give a local interaction $U/t \approx 10$, while $V \approx U/4$. At half-filling, this value of $U/t$ is large enough to put the system in the strong coupling limit with $\alpha \approx 1$, leading to an effective local interaction $\tilde{U}/t=(U-V)/t= 7.5$. If, however, as our data suggests, it is possible that $\alpha \approx 1.5$ for the doped case due to additional trionic screening channel, then the appropriate interaction for Hubbard model studies of the doped system would instead be roughly $\tilde{U}/t \approx 6.25$, which is a substantial reduction.

\begin{acknowledgments}
This work has been performed within the research program of the DFG Research Training Group \textit{Quantum Mechanical Materials Modeling} (QM$^3$) (Project P3). Furthermore, the work has been funded by the Cluster of Excellence ``Advanced Imaging of Matter'' of the Deutsche Forschungsgemeinschaft (DFG) - EXC 2056. 
EvL acknowledges support from the Swedish Research Council (Vetenskapsrådet, VR) under grant 2022-03090 and by eSSENCE, a strategic research area for e-Science, grant number eSSENCE@LU 9:1. G.N.G. acknowledges the Wallenberg Center for Quantum Technology (WACQT) for financial support via the EDU-WACQT program funded by Marianne and Marcus Wallenberg Foundation.
We thank Emanuel Gull for providing the CT-AUX code, and EK thanks Xinyang Dong for help at properly compiling it.
The authors gratefully acknowledge the computing time made available to them on
the high-performance computer Lise at the NHR Center ZIB (HLRN at the time). These
centers are jointly supported by the Federal Ministry of Education and Research
and the state governments participating in the NHR.
Further computations were performed using resources provided by LUNARC, The Centre for Scientific and Technical Computing at Lund University through projects 
LU 2023/2-37, 2024/2-24 and 2023/17-14.
\end{acknowledgments}

\appendix
\section{Computational details}
\label{sec:computational}

\begin{figure}[htp!]
	\centering
	\includegraphics[width=0.45\textwidth]{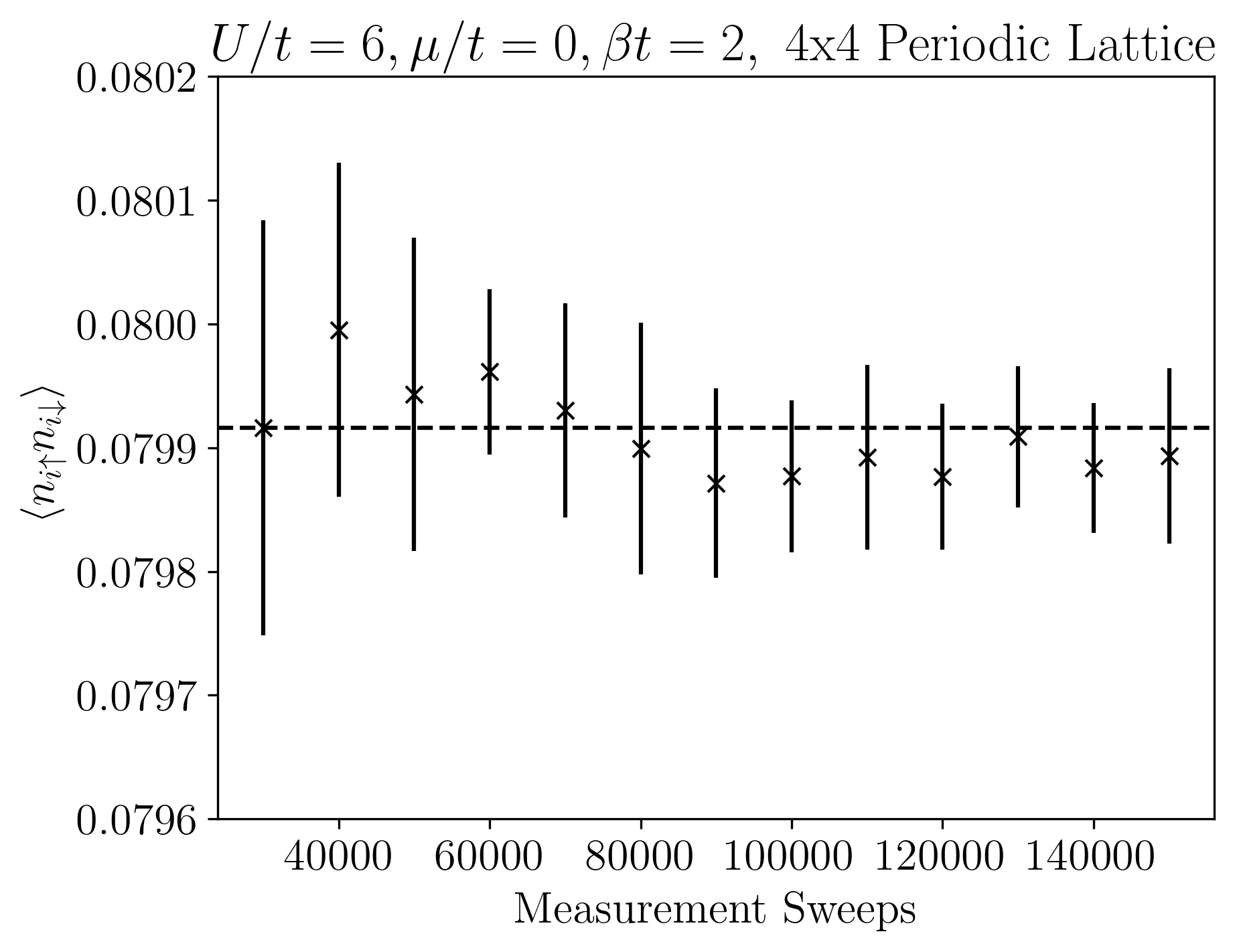}
	\caption{DQMC, QUEST code: Result for the double occupancy $\langle n_{i\uparrow} n_{i\downarrow} \rangle$ with error bars for different numbers of measurement sweeps.}\label{pic:DO_DQMC}
\end{figure}

\begin{figure}
	\centering
	\includegraphics[width=\columnwidth]{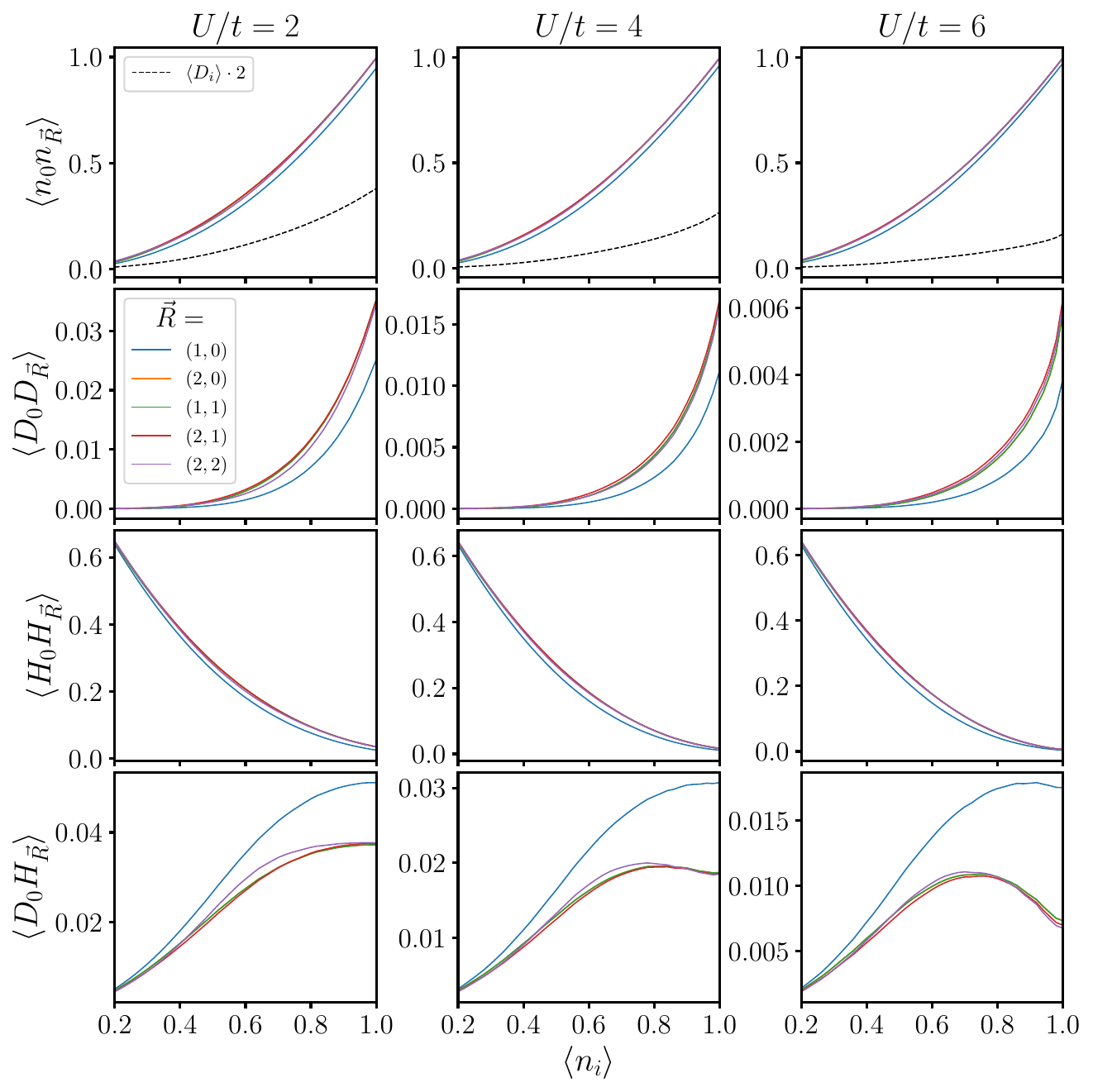}
	\caption{
 Spatial charge correlations (same as in Fig \ref{fig:4particle}, but for $U/t = 2,4,6$) in the Hubbard model, DQMC on a $4\times 4$ periodic cluster at $\beta t=2$.}
 \label{fig:4particle_U_246}
\end{figure}

In order to obtain the screening factor $\alpha (\tilde{U})$ (see Fig ~\ref{pic:AlphaResults}~\ref{pic:BetaDependence:alpha}), we solved the Hubbard Hamiltonian $\tilde{H} (H_{\text{Hub}})$, Eq.~\eqref{eq:hubbard}, on an equidistant 41x41 data grid ($\tilde{U}/t \in [0,8]$, $\tilde{\mu} / t \in [-8,0]$) for a temperature of $\beta t = 2$ within two different approaches: Determinantal Quantum Monte Carlo (DQMC) and the Dynamical Cluster approximation (DCA).
\subsubsection{DQMC}
The DQMC simulations have been performed with the QUEST code \cite{QUESTLink} on a 4x4 periodic lattice. The systematic error from the Trotter-Suzuki decomposition ($\mathcal{O}(\Delta \tau^2)$) can be minimized by choosing an appropriately small Trotter step ($\Delta \tau \sim \sqrt{0.125/\tilde{U}}$)~\cite{DQMCIntroDosSantos}. For the highest value of $\tilde{U}/t = 8$ considered here, this estimate would lead to $\Delta \tau \sim 0.125$. In order to keep the systematic error at a minimum, we choose $\Delta \tau = 0.05$ for the big simulations from which $\alpha$ is obtained, and $\Delta \tau = 0.02$ for the calculations of the four-particle correlators.

For each data point, the simulation is run with 10000 warmup sweeps and 30000 measurement sweeps. Since the system thermalizes quickly at the temperature considered, rather short simulations already yield appropriate results. As an example, Fig.~\ref{pic:DO_DQMC} shows, for a specific data point, that the correlation functions (here: the double occupancy) obtained from 30000 measurement sweeps are within error bars of simulations which are 5 times longer. Thus, for the purpose of evaluating accurate derivatives, a dense parameter grid combined with filtering is the more important aspect compared to the simulation time within QUEST. 

\subsubsection{DCA}

\begin{figure*}
	\centering
	\includegraphics[width=0.9\textwidth]{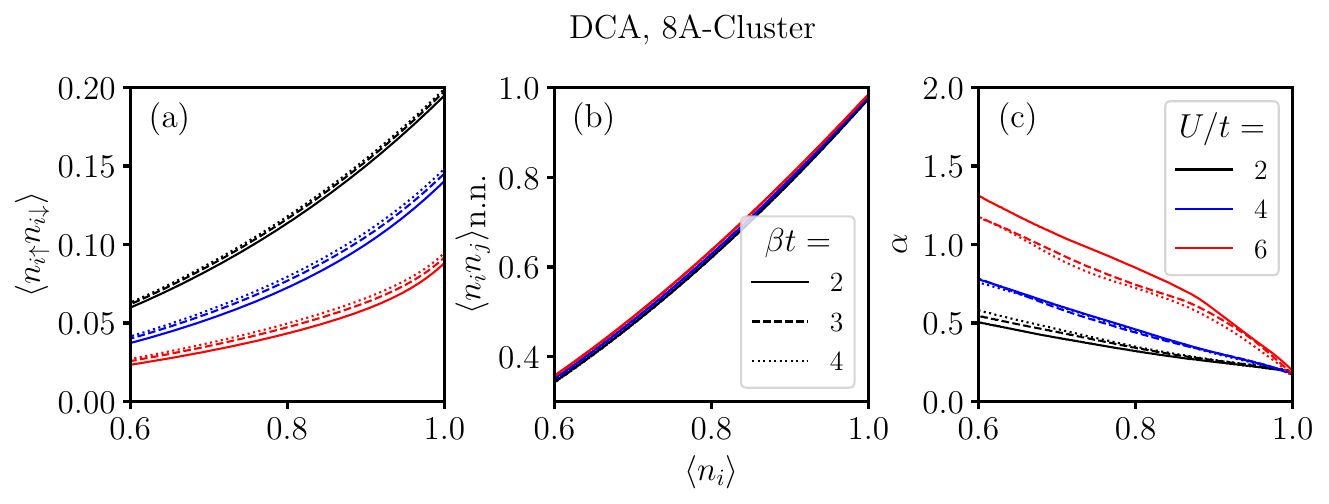}
	\caption{Double occupancy, next-neighbor density-density correlation and the resulting screening factor $\alpha$ as a function of interaction strength $U$, inverse temperature $\beta$ and filling $\langle n_i \rangle$, obtained using a 8-Site Cluster in DCA, c.f. Fig.~\ref{pic:BetaDependence:alpha} in the main text.}
 \label{pic:BetaDependence}
\end{figure*}

\begin{figure}[htp!]
	\centering
	\includegraphics[width=0.475\textwidth]{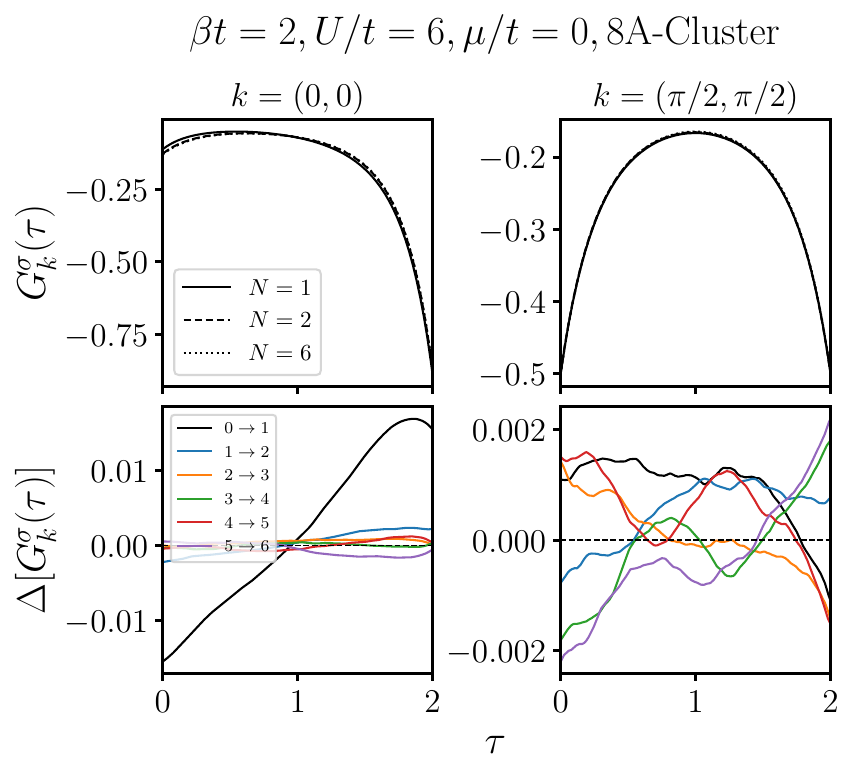}
	\caption{DCA, CT-AUX solver: Top pictures: Spin-averaged Green's function $G_k^{\sigma}(\tau)$ at different iteration steps within the self-consistency cycle. Bottom: Differences $\Delta (G_k^{\sigma}(\tau))$ between Green's functions at different steps.}\label{pic:GTAU_DCA}
\end{figure}

For the DCA approach, we use the CT-AUX code of Ref.~\cite{CTAUXPaperEmanuel}. Here, we find the necessary correlation functions by solving dynamical clusters with 8 and 16 sites. The simulations were performed in a timed manner, i.e. 25 minutes runtime (8 CPUs) for each data point and iteration step respectively. Within the self-consistency cycle, we allow for up to 8 iterations, which is, at the given temperature, more than sufficient. Fig.~\ref{pic:GTAU_DCA} illustrates for a single data point (8-site cluster), at 2 $k$-points, how the Matsubara Green's function behaves from iteration to iteration. The top left and right plots show the spin-averaged $G_k^{\sigma}(\tau)$ after iteration steps $N$, while the bottom pictures show the difference in the Green's function between iterations. In this case, already at 2 steps within the cycle, further iterations no longer cause significant changes.

\subsubsection{Filtering and Interpolation}
The central quantity presented within this work, the screening factor $\alpha (\tilde{U})$, depends on derivatives of $\tilde{U}$-dependent correlation functions. Such derivatives are difficult to obtain from noisy data, which is inherent in any Monte Carlo method. As such, proper filtering of our data is necessary.

Here, similar to previous works~\cite{MalteThermodynamicsMITExtendedHubbard,EdinHubbardHeisenberg}, we make use of a two-dimensional Savitzky-Golay filter \cite{SavitzkyGolay} which, in a box width of $w_{\tilde{U}}$ and $w_{\tilde{mu}}$, fits a two-dimensional polynomial of the following form to the data:
\begin{equation*}
	p\left( \tilde{U},\tilde{\mu} \right) = \sum_{nm}^{NM} c_{nm} \tilde{U}^n \tilde{\mu}^m
\end{equation*}
The polynomials are of third order, and the box widths are set to $w = 1.5$ for the DQMC and the 8-site DCA data, while it is set to $w = 2.0$ for the 16-site DCA data due to higher noise. Additionally, data which is close to the original starting point $(\tilde{U}_0,\tilde{\mu}_0)$, is given additional weight through a tricubic weighting function $(1 - d^3)^3$ where the distance $d$ is defined as $d = \text{max} \left\lbrace |\tilde{U} - \tilde{U}_0| / w_{\tilde{U}}, |\tilde{\mu} - \tilde{\mu}_0| / w_{\tilde{\mu}} \right\rbrace$.

After filtering the data, we use two-dimensional cubic interpolation to extend our parameter grid from 41x41 to 401x401, which makes it more convenient to evaluate derivatives.

Within the derivation for Eq.(\ref{eq:AlphaDefinition}), we chose $\tilde{\mu}$ as a function of $\tilde{U}$ in order to obtain a fixed filling, i.e. $\langle n_i \rangle_{\tilde{H}} = \text{const.}$. As such, as the last step before evaluating derivatives of correlation functions and thus, $\alpha (\tilde{U})$, we use the fillings $\langle n_i \rangle$ obtained from the simulations to transform the $\tilde{\mu}$-axis to an $\langle n_i \rangle$-axis.


\section{Exact Diagonalization}
\label{sec:ED}

\begin{figure}
	\centering
	\includegraphics[width=\columnwidth]{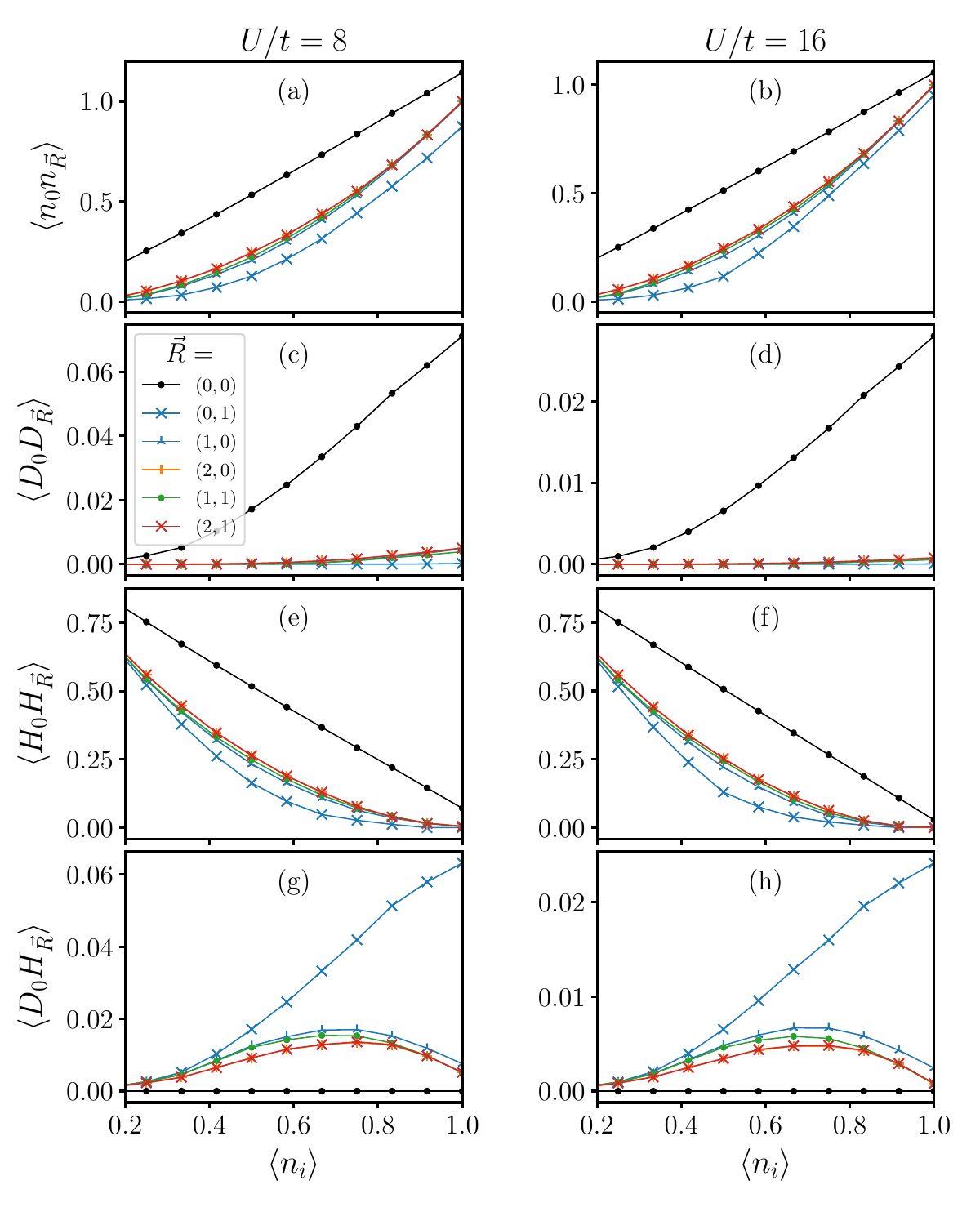}
	\caption{
 Spatial charge correlations in the Hubbard model, ED on a $4\times 2$ periodic cluster at $T=0$. (a-b) Density-density, (c-d) doublon-doublon, (e-f) hole-hole (g-h) doublon-hole.
 }
 \label{fig:4particle:ED}
\end{figure}

To verify our numerical results, we have also performed exact diagonalization (ED) simulations of a $4\times 2$ ribbon with periodic boundary conditions in both directions (i.e., a donut), using EDLib~\cite{Iskakov18}. We extract the relevant correlation functions from the ground state wavefunctions in a particular sectors $(N_\up,N_\dn)$. The Exact Diagonalization is complementary to the QMC in several ways, it is performed at $T=0$ directly and with fixed particle number, instead of the grand canonical ensemble used in QMC. It is sign-problem free and can be done at any $U$. On the other hand, the scaling of the computational effort with system size is much worse than in QMC.

Figure~\ref{fig:4particle:ED} shows the results of the ED, which can be compared with Fig.~\ref{fig:4particle}. Note that the $4\times 2$ geometry with periodic boundary conditions in both directions leads to a difference between the $\vec{R}=(0,1)$ and $\vec{R}=(1,0)$ correlation functions, since hopping along the short direction is enhanced. Thus, to optimize the kinetic energy, doublon-holon excitons preferentially align along this axis. In this way, the broken rotational symmetry of the system actually provides strong evidence for the kinetic energy as the mechanism for the binding. A similar mechanism has been discussed in $t-J$-bilayers~\cite{Bohrdt22}.

Apart from the special $\vec{R}=(0,1)$ in the $4\times 2$ cluster, the results at $U/t=8$ are qualitatively similar and even quantitatively close to Fig.~\ref{fig:4particle}, even though there is a difference in cluster size and statistical ensemble. This shows that generic mechanisms of the Hubbard model are responsible for these effects. In ED, we can also increase $U$ further, to $U/t=16$. The density-dependence of the curves does not change, but the overall magnitude of the doublon correlators decreases substantially due to the decrease in the number of doublons. This shows that the $U/t=8$ QMC results in the main text already give a good impression of what happens at strong coupling. 

\bibliography{paper.bib}
\end{document}